\documentclass[aps,prb,twocolumn,showpacs,amsmath,floatfix]{revtex4}
\usepackage{graphicx}
\usepackage{dcolumn}
\usepackage{multirow}

\begin{document}
\title{Structural and electronic origin of the magnetic structures in hexagonal LuFeO$_3$}

\author{Hongwei Wang$^{1,2}$, Igor V. Solovyev$^{3}$, Wenbin Wang$^{4}$, Xiao Wang$^{5}$, 
Philip J. Ryan$^{6}$,  David J. Keavney$^{6}$, Jong-Woo Kim$^{6}$, Thomas Z. Ward$^{7}$, Leyi Zhu$^{8}$,
Jian Shen$^{4}$, X. M. Cheng$^{5}$,  Lixin He$^{2}$,  Xiaoshan Xu$^{5,7,9,*}$, and Xifan Wu$^{2, *}$}

\affiliation{$^{1}$Department of Physics,  
Temple University, Philadelphia, Pennsylvania 19122, USA}
\affiliation{$^{2}$Key Laboratory of Quantum Information,
University of Science and Technology of China, Hefei, Anhui, 230026, People's
Republic of China}
\affiliation{$^{3}$Computational Materials Science Unit, National Institute for Materials Science, 1-2-1 Sengen, Tsukuba 305-0047, Japan}
\affiliation{$^{4}$Department of Physics, Fudan University, Shanghai 200433, China}
\affiliation{$^{5}$Department of Physics, Bryn Mawr College, Bryn Mawr, PA 19010, USA}
\affiliation{$^{6}$Advanced Photon Source, Argonne National Laboratory, Argonne, IL 60439, USA}
\affiliation{$^{7}$Materials Science and Technology Division, Oak Ridge National Laboratory, Oak Ridge, TN 37831, USA}
\affiliation{$^{8}$Materials Science Division, Argonne National Laboratory, Argonne, IL 60439, USA}
\affiliation{$^{9}$Department of Physics and Astronomy, University of Nebraska-Lincoln, NE 68588, USA}


\begin{abstract}
Using combined theoretical and experimental approaches, we studied the structural and electronic origin of the magnetic structure in hexagonal LuFeO$_3$.
Besides showing the strong exchange coupling that is consistent with the high magnetic ordering temperature, 
the previously observed spin reorientation transition is explained by the theoretically calculated magnetic phase diagram.
The structural origin of this spin reorientation that is responsible for the appearance of spontaneous magnetization, 
is identified by theory and verified by x-ray diffraction and absorption experiments.
\end{abstract}
\maketitle

\section{Introduction}
While the ferroelectricity in materials is naturally connected to structural distortions 
that breaks the spatial inversion symmetry~\cite{Stengle_Nature_physics, Fu_Nature}, the relation between spontaneous magnetization 
and structure is not obvious because no spatial symmetry is broken by ferromagnetism (FM).
Nevertheless, magnetic orderings in a material are tied to the structure, and the ties are particularly 
important in multiferroic materials~\cite{2008Lee} in which structural distortions may mediate 
couplings between ferroelectricity and ferromagnetism or even 
generate ferroelectric ferromagnets which are extremely rare~\cite{2010June}.

\begin{figure}
\includegraphics[width=3.0in]{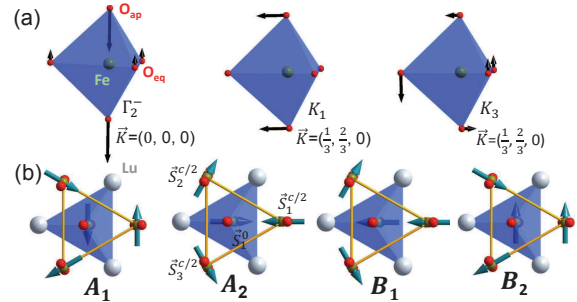}
\caption{\label{fig_phonon_mag} 
(Color online.) 
(a) Displacement patterns of the FeO$_5$ local environment (trigonal bipyramid) in the three phonon modes that freeze in the P6$_3$/mmc $\rightarrow$ P6$_3$cm structural transition in hexagonal ferrites (h-RFeO$_3$). 
 The arrows indicate the relative displacement of the atoms.
  $\vec{K}$ is the wave vector of the modes in the reciprocal space of the P6$_3$/mmc structure.
(b) Four independent spin structures ($A_1$, $A_2$, $B_1$, and $B_2$) of the 120-degree magnetic orders viewed along the $c$ axis.
 The arrows indicate the spins ($\vec{S}^{Z_{Fe}}_i$) on the Fe sites.
 The Fe sites shown in the polyhedra are in the $Z_{Fe}=0$ layer while all the other Fe sites are in the $Z_{Fe}=c/2$ layers.
 In the $B$ ($A$) phase, $\vec{S}^{0}_1$ is parallel (antiparallel) to $\vec{S}^{c/2}_1$. 
 }
\end{figure}

The recently discovered room temperature multiferroic, i.e. hexagonal LuFeO$_3$ (hLFO)~\cite{Xu_PRL_2013},
provides an rare case multiferroic material in which spontaneous electric and magnetic polarizations coexist.
On one hand, ferroelectricity appears below T$_{\rm C}$=1050 K resulting from a 
P6$_3$/mmc $\rightarrow$ P6$_3$cm structure distortion, which can be decomposed 
in terms of three phonon modes (Fig.~\ref{fig_phonon_mag}(a))~\cite{Xu_PRL_2013,2010Magome}.
On the other hand, spin frustration in hLFO presents rich magnetic phases~\cite{2000Munoz}.
Intriguingly, below the Ne\'el temperature T$_{\rm N}=440$ K,  
magnetic order in hLFO transits again from $B_2$ to $A_2$ (Fig. \ref{fig_phonon_mag}(b)) 
at T$_{\rm R}$=130 K~\cite{Xu_PRL_2013} by a spin reorientation (SR), resulting in a weak ferromagnetism due to the
Dzyaloshinskii-Moriya and single-ion anisotropy mechanism~\cite{Dzyaloshinsky, Moriya, Vanderbilt_TMO, Hong_PRB, 2011Akbashev}.
Similar to hexagonal YMnO$_3$, the $K_3$ phonon is believed to be the driving force that 
induces the instability of $\Gamma_2^-$ that is responsible for the ferroelectricity~\cite{Fennie_LFO, Wu_LFO, Rabe_YMO}.
However, the origin of the SR is still elusive. 
Since the SR is the direct cause of spontaneous magnetization, 
elucidating the origin may provide a way to effectively tune T$_{\rm R}$, 
or even a novel route for realizing a coexistence of spontaneous electric and 
magnetic polarizations above room temperature~\cite{Physics_Today, BFO_Science, BFO_Theory}.

 Previous studies in hexagonal manganites (h-RMO, isomorphic to hLFO) indicate rich magnetic phases due to the SR that is strongly coupled to the crystal structure ~\cite{2007Lancaster,2008Lee,2009Fabreges}. 
 However, the multiple degrees of freedom involved (spin and orbital degrees of freedom of the electrons and the lattice) complicate the problem in h-RMO ~\cite{Brown}. 
 The complexity may be reduced in h-LFO, in which Fe$^{3+}$ can be considered a spin-only ion with nearly spherical $3d^5$ electronic configuration.
 Therefore, a better understanding on the SR in hLFO is possible, particularly in terms of the phonon modes (Fig. \ref{fig_phonon_mag}(b)); it may also be an important step in understanding the more complex SR in h-RMO~\cite{Brown}, in which the single-ion anisotropy is expected to play a more important role.

To address the above issues,  we perform 
combined theoretical and experimental studies of the exchange 
interactions and its couplings to the structural instabilities in
hLFO. We apply an extended Kugel-Khomskii (KK) model for
superexchange (SE) interactions~\cite{Model} based
on localized Wannier functions (LWFs)~\cite{MLWF, Selloni_MLWF}.
While the antiferromagnetic (AFM) exchange coupling is dominated by the intralayer superexchange, 
the model clearly shows that the singly occupied 
$d_{z^2}$ orbital in hLFO greatly increases the exchange coupling 
compared with the empty $d_{z^2}$ in LuMnO$_3$ (LMO).
The interlayer exchange, although much weaker in magnitude, is key to
the SR. Our first-principles calculations show that
SR is strongly coupled to the $K_1$ phonon mode 
and only weakly dependent on $K_3$ mode. 
Our theory indicates that the atomic displacements of $K_1$ mode is responsible for the SR.
This scenario is then confirmed by our x-ray diffraction and x-ray absorption experiments.

\section{Computational methods and experimental techniques}
Our extended KK model~\cite{Model, Support} is built on the basis of LWFs generated from 
density functional theory (DFT) calculations.
The screened Coulomb interactions between LWFs are computed in the constrained random-phase
approximation~\cite{Igor_PRB, Igor_review, Support}.
The calculations of spin phonon coupling is performed within DFT+U scheme~\cite{VASP, Support}. 
We have adopted the four-state method~\cite{Xiang_PRB_2011} 
in computing the exchange coupling strengths.
hLFO films (50 nm thick) were grown on  Al$_2$O$_3$ (0001) substrates with and without a (30 nm) Pt buffer layer using pulsed laser deposition. 
 The x-ray diffraction (XRD) and x-ray absorption spectroscopy (XAS)  measurements were carried out in 6-ID-B beam line on the h-LuFeO$_3$/Al$_2$O$_3$ film and in 4-ID-C beam line on the h-LuFeO$_3$/Pt/Al$_2$O$_3$ film respectively at the Advanced Photon Source  at various temperature. 

\section{Results and discussion}
In hexagonal ferrites, the exchange interaction between the Fe sites can be written as
\begin{eqnarray}
H_{ex} & = & H_{ex}^{a-b}+H_{ex}^{c}
\label{eq:Hex}
\end{eqnarray}
where  $H_{ex}^{a-b}$ is the intra-layer exchange interaction
and $H_{ex}^{c}$ is the inter-layer exchange interaction considering only the nearest neighbors.

\begin{figure}
\includegraphics[width=3.0in]{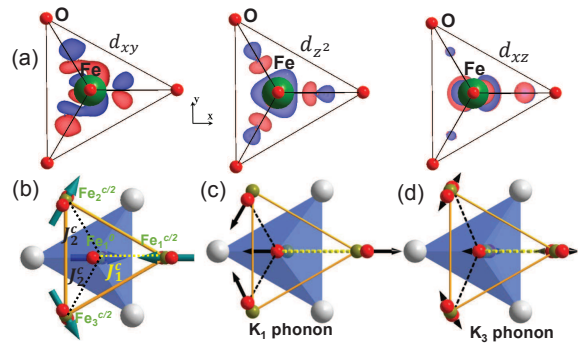}
\caption{\label{fig1} (Color online.) (a) Representative $d_{xy}$,  $d_{z^2}$, and $d_{xz}$-like
LWFs viewed from [001] direction. (b)illustrations of two independent SSE paths $J_1^c$
and $J_2^c$  between Fe$_0$ at $z/c=0$ and three neighboring iron ions Fe$_1$, Fe$_2$, and Fe$_3$
at $z/c=1/2$ (c)atomic displacements of $K_1$ phonon mode,
(d)atomic displacements of $K_3$ phonon mode, viewed from [001] direction. }
\label{figure1}
\end{figure}

As shown in Fig. \ref{fig1}, the intra-layer SE interaction  $H_{ex}^{a-b}=\sum\limits_{{ i},j,Z_{Fe}} { {\mathcal J}_{i,j}^{a-b}\vec{S}_i^{Z_{Fe}} \cdot \vec{S}_j^{Z_{Fe}}}$ between two nearest neighbor (NN) Fe atoms at site $i$ and $j$
are mediated by corner sharing oxygen atoms.
In order to elucidate the electronic structural origin, we employ
the extended KK model and the SE coupling can be expressed as 
%
\begin{eqnarray}
{\mathcal J}_{i,j}^{a-b} & =&  \sum_{\alpha, \alpha '}J_{\alpha, \alpha '}^{{\rm AFM}} + \sum_{\alpha, \beta}J_{\alpha , \beta}^{{\rm FM}} 
\label{eq:CE_general}
\end{eqnarray}
%
The first term in Eq. (\ref{eq:CE_general}) describes 
the AFM-type coupling resulting from virtual hopping processes between two half-filled $d$ bands; while the second term 
depicts the competing FM-type coupling from hoppings from half-filled $d$ orbital ($\alpha$)  to empty ones ($\beta$)~\cite{Support}. 
The computed  individual exchange interaction  as well as the overall SE coupling $J_{\rm MOD}^{a-b}$ for both hLFO and hLMO 
are presented in Table~\ref{Couplings}. The total exchange coupling $J_{\rm DFT}^{a-b}$ from the direct 
fit of the total DFT energies are also shown.
\begin{table}[hb]
\begin{center}
 \caption{Individual and total intralayer exchange interaction(meV) in both hLFO and hLMO~\cite{Support}.}
\begin{tabular}{c|rrrrrrrr}
  \hline
   \hline 
       $J_{\alpha, \alpha ' (\beta)}$    &       & $d_{xy}$ &  $d_{x^{2}-y^{2}}$       & $d_{z^{2}}$  & $d_{xz}$ & $d_{yz}$ 
& $J_{\rm MOD}^{ab}$ & $J_{\rm DFT}^{ab}$ \\
  \hline 
  \multirow{5}{*}{{\rm hLFO}}
  & $d_{xy}$               & 9.49 & 3.65 & 3.20 & 1.14 & 0.47 &\multirow{5}{*}{{45.2}}& \multirow{5}{*}{{49.7}}\\
   & $d_{x^{2}-y^{2}}$   & 0.68 & 9.90 & 0.88 & 1.04 & 0.05  \\
 & $d_{z^{2}}$    & 1.25 & 5.05 & 3.58 & 1.25 & 0.27  \\
  & $d_{xz}$               & 0.56 & 0.09 & 0.14 & 0.01 & 0.06  \\
 & $d_{yz}$               & 0.32 & 1.00 & 0.65 & 0.37 & 0.06  \\
  \hline
   
    \hline
\multirow{5}{*}{\rm hLMO}
 &$d_{xy}$               & 10.15  & 5.81 & $-$0.64  & 1.17  & 0.78 &\multirow{5}{*}{{29.3}}& \multirow{5}{*}{{30.7}}  \\
   & $d_{x^{2}-y^{2}}$   & 1.28 & 10.9 & $-$2.71 & 0.85 & 0.12  \\
 & $d_{z^{2}}$    & - & - & - & - & -  \\
  & $d_{xz}$               & 0.41 & 0.17 & $-$0.43 & 0.01 & 0.08  \\
 & $d_{yz}$               & 0.39 & 0.85 & $-$0.19 & 0.31 & 0.04  \\
  \hline
\end{tabular}
\label{Couplings}
\end{center}
\end{table}

According to the local environment(Fig. \ref{fig_phonon_mag}), 
the $3d$ orbitals in Fe and Mn are split into $e''(xz, yz)$, $e'(x^2-y^2, xy)$, and $a_1'(z^2)$ levels by the crystal field, with the increasing energy respectively~\cite{Xu_APL_2012,2007Cho}.
It can be seen that the largest SE interactions are contributed by the 
diagonal hopping processes involving $d$ orbitals of $e'$ symmetry.
This is consistent with the physical expectation that SE is of intralayer nature
while $d_{xy}$ and $d_{x^2-y^2}$ are the only $d$ orbitals lying mostly inside the $ab$ plane.
Centered on the magnetic ions, these $d$-like 
LWFs are also connected with its first neighboring magnetic atoms through the hybridization 
with the shared oxygen atoms on the bipyramids.
As a result, a strong oxygen $p$ character is found on the lobe of the LWFs, 
pointing to each of the three neighboring oxygen atoms.
Considering such $d$-like LWFs on the hexagonal lattices,
a large AFM hopping integral is thus expected
along the path of Fe(Mn)-O-Fe(Mn)~\cite{Anderson}.
Based on the same orbital symmetry argument, it can be easily seen that 
the diagonal hopping is relatively smaller for $a_1'(z^2)$ character and almost
zero for $e''$ character. This is because $d_{z^2}$ and $d_{xz}$($d_{yz}$)
require that the main orbital lobe to be located along $z$ or within the $xz$($yz$)
plane which make the hopping integrals much smaller.

Strikingly, the SE interactions also show fundamental difference between the two materials.
In hLFO(Fe$^{3+}$:$3d^54s^0$), the $d_{z^2}$ orbital of $a_1'$ symmetry is singly occupied
and SE interactions can only be of AFM types. However, 
$d_{z^2}$ is empty in hLMO (Mn$^{3+}$:$3d^44s^0$), SE interactions
are thus composed of competing AFM and FM types and
the coupling strength is further reduced by the forbidden 
hopping involving the empty $d_{z^2}$. Thus, a significantly 
larger AFM coupling energy is observed in hLFO. This 
is consistent with the higher Ne\'el temperature in hLFO observed in experiment
in addition to the larger spin on the Fe site.

\begin{figure}
\includegraphics[width=3.39in]{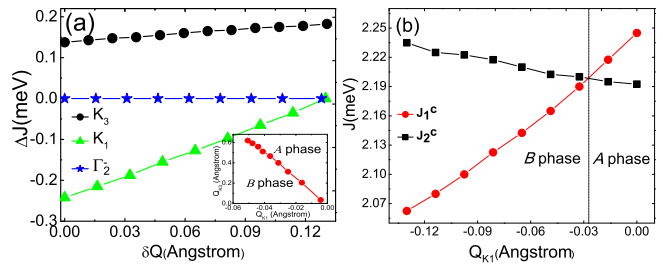}
\caption{\label{fig3} (Color online.)   (a) $\Delta J(\delta Q)=J(Q)-J(Q_0)$ for each individual $K_1$, $K_3$, and $\Gamma_2^{-}$ phonon mode,
 where $\delta Q=Q-Q_0$ and $Q_0$ is the value at 300 K, 
 while the other two phonon displacements are kept as zero.
Insert: theoretical phase diagram as functions of mode amplitudes of $K_1$ and $K_3$.
 (b) $J_1^c$ and $J_2^c$ as functions of $Q_{K_1}$,  
 while $Q_{K_3}$ and $Q_{\Gamma_2^-}$ are fixed at the experimental values~\cite{Support}.}
\label{spinphonon_coupling}
\end{figure}

Having established the electronic origin of the large intralayer
exchange coupling, we now focus on the interlayer exchange coupling
$H_{ex}^c$=$\sum\limits_{i,j,Z_{Fe}} {{\mathcal J}_{i,j}^{c}\vec{S}_i^{Z_{Fe}} \cdot \vec{S}_j^{Z_{Fe}+\frac{c}{2}}}$
in hLFO. This is the key to understanding the mechanism
of SR and weak FM moment below T$_{\rm R}$~\cite{Xu_PRL_2013}.
Compared to the SE nature of intralayer exchange, 
the interlayer Fe ions are coupled by the super super exchange interaction (SSE)~\cite{SSE}, 
in which one Fe atom at $Z_{Fe}=0$ is in exchange interaction with three first neighbor Fe atoms at $Z_{Fe}=c/2$ 
mediated by two apical oxygen atoms (O$_{ap}$).
Due to the P6$_{3}$cm structure in Fig. \ref{spinphonon_coupling} (b), the three SSE paths can be further simplified by
two independent SSE coupling strengths:
 $J_1^c$ through Fe$_1^0$-O-$\cdot\cdot\cdot$-O-Fe$_1^\frac{c}{2}$ and $J_2^c$
through Fe$_1^0$-O-$\cdot\cdot\cdot$-O-Fe$_2^\frac{c}{2}$ respectively. 
As a result, the $H_{ex}^c$ spin Hamiltonian in Eq. (\ref{eq:Hex}) can 
be rewritten as 
$H_{ex}^c=\sum\limits_{i,Z_{Fe}} {({J_1}^{c}-{J_2}^{c})\vec{S}_i^{Z_{Fe}} \cdot \vec{S}_i^{Z_{Fe}+\frac{c}{2}}}$.
Obviously, the sign of $\Delta J = J_1^c - J_2^c$ determines the preferred alignment between 
$\vec{S}_i^{Z_{Fe}}$ and $\vec{S}_i^{Z_{Fe}+\frac{c}{2}}$:
parallel ($B$ phase) if $\Delta J < 0$; 
antiparallel ($A$ phase) if $\Delta J > 0$;
no alignment if $\Delta J=0$, which is the case for P6$_3$/mmc structure.

Since the non-zero $\Delta J$ comes from the structural distortion (P6$_3$/mmc $\rightarrow$ P6$_3$cm), 
the low temperature spin reorientation must have a structural origin.
Here we investigate the dependence of $\Delta J$ on the three phonon modes $K_1$, $K_3$, and $\Gamma_2^{-}$ 
that are responsible for the structural distortion~\cite{Support}. 
We use DFT to calculate the $\Delta J$ as functions of phonon mode displacements ($Q_p$, where $p$=$K_1$, $K_3$, and $\Gamma_2^{-}$)
and the results are shown in Fig. \ref{spinphonon_coupling}(a). It can be seen that $\Delta J$ depends on the displacement of each phonon mode rather differently.

\begin{figure*}
\includegraphics[width=6.0in]{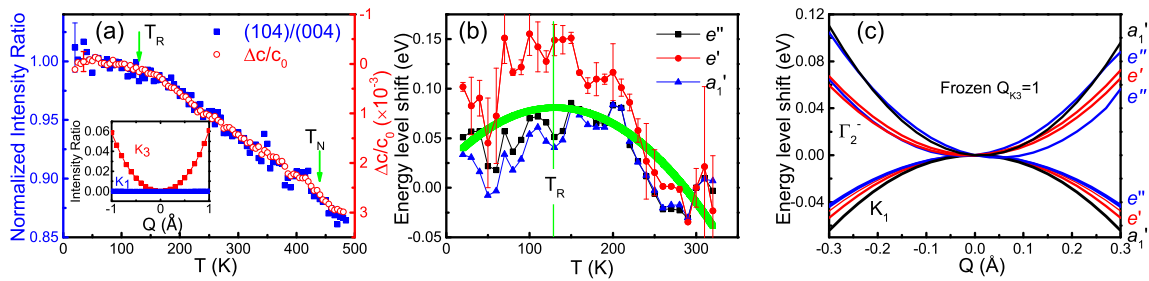}
\caption{(Color online.) Structural changes indicated by the XRD and XAS. 
(a) The XRD intensity ratio between the (104) and (004) peaks (normalized to the 30 K value) and the change of lattice constant $c$ (with respect to the 30 K value) as functions of the temperature (the representative error bars are shown).
Inset: the simulated intensity ratio between the (104) and (004) peaks as functions of phonon displacements.
(b) The change of Fe-3d crystal field levels (relative to the 300 K values) as functions of the temperature (the error bars for $e'$ levels are shown as examples); 
 the bold line is a guide to the eye to highlight the common peak-like feature.
(c) Simulated change of Fe-3d levels as functions of phonon displacements~\cite{Support}.
\label{Fig_xray}}
\end{figure*}

Clearly, $K_1$ phonon mode has the largest effect on SR. 
This can be identified by the steepest slope of $\Delta J$ when $K_1$ is increased perturbatively, 
yielding a linear coefficient $\frac{\delta \Delta J}{\delta Q_{K_1}} \sim 1.9 {\rm meV}/$\AA.
This suggests a strong tendency of $K_1$ in driving hLFO from $B$ phase ($\Delta J < 0$) 
into $A$ phase ($\Delta J > 0$).
Indeed this is also consistent with the physical expectation of atomic displacements under the $K_1$ mode.
$K_1$ phonon is a Brillouin Zone (BZ) boundary mode and is of pure in-plane nature.
The atomic displacements of $K_1$ phonon mostly involve the O$_{ap}$ of FeO$_5$ (Fig. \ref{fig_phonon_mag}).
As shown in Fig. \ref{figure1}(c), the effects of the $K_1$ are: 
O$_{ap}$ of Fe$_1^0$ moves away from that of Fe$_1^\frac{c}{2}$, causing $J_1^c$ to decrease;
O$_{ap}$ of Fe$_1^0$ moves closer to that of Fe$_2^\frac{c}{2}$ and Fe$_3^\frac{c}{2}$, causing $J_2^c$ to increase.
As a result, K$_1$ is strongly coupled to the $\Delta J$. 

The $K_3$ phonon mode can be described by the rotation of FeO$_5$ (Fig. \ref{fig_phonon_mag}(a)) also located at BZ boundary.
The atomic displacements of $K_3$ mode include all the O$_{ap}$ of the FeO$_5$.
However, due to its rotational nature, the atomic displacement of the O$_{ap}$ alternate their directions along $c$ as shown in Fig \ref{figure1}(d).
As a result, the overall length of $J_1^c$ and $J_2^c$ paths are barely 
changed except that the Fe atom is slightly moved away from its equilibrium positions in P$6_{3}/$mmc symmetry.
Compared with the direct tunability of $\Delta J$ by $K_1$ mode, 
the $K_3$ phonon is expected to be a second order effect in SR.
Indeed, our DFT calculation predicts a much weaker variation of $\Delta J$ with
increased $K_3$ phonon mode amplitude, in which the linear 
coefficient $\frac{\delta \Delta J}{\delta Q_{K_3}} \sim 0.3 {\rm meV}/$\AA~is 
about one order of magnitude smaller than that of $K_1$.
Similar to that of $K_1$ mode, the slope is also positive, favoring the SR from $B$ to $A$ phase.

Finally, we focus on the coupling between $\Gamma_2^{-}$ and $\Delta J$.
$\Gamma_2^{-}$ is ferroelectric phonon mode at zone center. The atomic displacements of this mode
involve all the Lu, O, and Fe atoms moving along $c$. However, the 
displacements of the two O$_{ap}$ of one bipyramid are exactly the same.
As a result, the SSE paths in $J_1^c$ and $J_2^c$ are 
changed uniformly. Not surprisingly, our theory predicts a zero dependence
of $\Delta J$ on $\Gamma_2^{-}$ mode amplitude. It indicates that this ferroelectric
distortion alone does not play any role in SR.

The significantly different coupling strengths of $\Delta J$ with the
phonon modes suggests the primary role of $K_1$ phonon mode in SR of hLFO.
Indeed, when $K_1$ mode is frozen into the experimental structural 
coordinates at T=300 K perturbatively,
$J_1^c$ and $J_2^c$ rapidly increases and decreases respectively and SR occurs
at the crossing point as shown in Fig. 3(b) separating the $B$ from $A$ phase. 
Below, we show that $Q_{K_3}$ saturates at close to T$_{\rm R}$, while $Q_{K_1}$ changes significantly from 300 K to 20 K using XRD and XAS measurements.

As shown in Fig.\ref{Fig_xray}(a), the temperature dependence of the normalized intensity ratio
between (104) and (004) peaks appears to saturate when temperature is lowered to T$_{\rm R}$.
 We attribute the saturation to the slow variation of ${K_3}$ phonon at low temperature, because $K_3$ is expected to have a dominant effect here, according to the simulated intensity ratio (Fig.\ref{Fig_xray}(a) inset)~\cite{Cullity1956}, while the zone center mode $\Gamma_2$ is expected to have no effect.
 The saturation of ${K_3}$ mode can be further confirmed by the temperature dependence of the lattice constant $c$ which follows closely to that of the intensity ratio, as shown in Fig.\ref{Fig_xray}(a). 
 The displacement of the $K_3$ mode includes a rotation of the FeO$_5$ trigonal bipyramid, which changes the shape of the unit cell by enlarging $a$ and reducing $c$~\cite{2000Munoz}; 
 the change of $c$ ($\Delta c$) is proportional to $\Delta Q_{K_3}$ for small change of $Q_{K_3}$.
 The matching temperature dependence in Fig.\ref{Fig_xray}(a) suggests that the change of $c$ is indeed caused by the $K_3$ mode which saturates at low temperature.

 XAS measurements suggest that the $K_1$ mode undergoes a gradual change at low temperature.
 Previously, we assigned the Fe-3d crystal levels using the XAS at room temperature ~\cite{Xu_APL_2012,Support}.
	As shown in Fig. \ref{Fig_xray}(b), the temperature dependences of the energy levels all show broad peak-like features with the maxima close to T$_{\rm R}$.
	The crystal field levels of Fe-3d are expected to be sensitive to the shape of FeO$_5$.
	As shown in Fig. \ref{fig_phonon_mag}(a), the $K_3$ mode causes a rotation of the FeO$_5$ while the $K_1$ or $\Gamma_2^-$ modes cause distortions of the FeO$_5$, so the energy-level shift observed in Fig. \ref{Fig_xray}(c) are most likely generated by the change of $Q_{K_1}$ or $Q_{\Gamma_2^-}$.
	Fig. \ref{Fig_xray}(c) shows a simulation~\cite{Support} of the energy level change of the crystal field levels as functions of $Q_{K_1}$ or $Q_{\Gamma_2^-}$ with respect to the value when all the mode displacements are zero.
	According to the simulation, the $K_1$ mode generates a maximum at $Q_{K_1}=0$ while $Q_{\Gamma_2^-}$ generates a minimum at $Q_{\Gamma_2^-}=0$; 
	this is because the $K_1$ mode move both O$_{ap}$ atoms away from the Fe sites and make the FeO$_5$ larger while the $\Gamma_2^-$ mode pushes one O$_{ap}$ atom close to Fe site.
	Comparing the simulation and the observation, we infer that the $K_1$ mode changes gradually when the temperature is lowered, in order to generate the maximum \cite{Support};
	this is consistent with the theoretical prediction in which $Q_{K_1}$ changes when the temperature is lowered and causes the transition from antiferromagnetism in B$_2$ phase to weak ferromagnetism in A$_2$ phase. 
 
\section{Conclusion} 
 Hence, the roles of all three structural distortions are elucidated in hLFO: 
 the instability of $K_3$ mode is the driving force of the P6$_3$/mmc $\rightarrow$ P6$_3$cm structural transition; 
 the improper ferroelectricity of $\Gamma_2^-$ mode is induced by frozen $K_3$ mode~\cite{Rabe_YMO,Fennie_LFO};
 the competing effect between $K_1$ and $K_3$ modes determines the magnetic ordering and drives the magnetic phase transition.
If the $K_1$ mode can be tuned by interface engineering~\cite{Rabe_PRL_2010, Dieguez_strain, Junquera_Nature}, 
the $T_{\rm R}$ can be increased, achieving the spontaneous electric and magnetic polarizations and their couplings at room temperature.

\section*{Acknowledgment}
This work is supported by Air Force Office of Scientific Research under award No. FA9550-13-1-0124 (X.Wu). 
Computational support is provided by the National Energy Research Scientific Computing Center and
 by the National Science Foundation through XSEDE resources provided by the XSEDE Science 
Gateways program (award No. TG-DMR120045). Research supported by the U.S. Department of Energy, Basic Energy Sciences, 
Materials Sciences and Engineering Division (T.Z.W., X.S.X.). 
We also acknowledge partial funding support from the National Basic Research Program of China (973 Program) 
under the grant No. 2011CB921801 (J.S.), the National Natural Science Funds of China Grant No. 11374275 (L.H.) 
and the US DOE Office of Basic Energy Sciences, the US DOE grant DE-SC0002136 (W.B.W.). Use of the Advanced Photon 
Source was supported by the U. S. Department of Energy, Office of Science, Office of Basic Energy Sciences, 
under Contract No. DE-AC02-06CH11357. X. M. Cheng acknowledges support from the National Science Foundation under Grant No. 1053854. 
 
X. Wu is grateful for the useful discussions with Andrei Malashevich, Craig Fennie, Weida Wu, and David Vanderbilt. 

$^*$ To whom correspondence should be addressed: xifanwu@temple.edu, and xiaoshan.xu@unl.edu.

\end{document}